\documentclass[letter]{aa} 
\usepackage{graphicx}
\usepackage{txfonts}
\usepackage{natbib}

\bibpunct{(}{)}{;}{a}{}{,}
\begin{document}
   \title{Invisible sunspots and rate of solar\\
 magnetic flux emergence\\[1em]}

   \author{S.~Dalla\inst{1} \and L.~Fletcher\inst{2} \and  N.A.~Walton\inst{3} }

   \authorrunning{Dalla et al.}

   \titlerunning{Invisible sunspots}

   \institute{Centre for Astrophysics, University of Central Lancashire,
              Preston PR1 2HE, UK. 
         \and
              Department of Physics and Astronomy,
               University of Glasgow, Glasgow G12 8QQ, UK.
         \and
              Institute of Astronomy,
              University of Cambridge, Cambridge CB3 OHA, UK.
           }

   \date{Received October 5, 2007}

\abstract{}
{We study the visibility of sunspots and its influence 
on observed values of sunspot region parameters.}
{We use Virtual Observatory tools provided by AstroGrid
to analyse a sample of 6862 sunspot regions. By studying
the distributions of locations where sunspots were first and
last observed on the solar disk, we derive the visibility
function of sunspots, the rate of magnetic flux emergence and the
ratio between the durations of growth and decay phases of solar
active regions.}
{We demonstrate that the visibility of small sunspots 
has a strong centre-to-limb variation, far larger than would be expected
from geometrical (projection) effects.
This results in a large number of young spots being 
invisible: 44\% of new regions emerging in the west of the Sun
go undetected.
For sunspot regions that are detected, large differences exist
between actual locations and times of flux emergence, and the apparent
ones derived from sunspot data. The duration of the growth phase of
solar regions has been, up to now, underestimated.
}
{} 
 

                 
   \maketitle
%

\section{Introduction}

The birth of a new spot on the solar disk indicates 
the emergence of magnetic flux through the photosphere, a
process which is key to the solar cycle \cite{Sol2003,Fis2000} and the
study of stellar magnetic dynamos. Sunspots also cause variations in 
the total solar irradiance, 
an important parameter in determining the Sun's influence on climate \cite{Fou2006}.
The presence of a sunspot is key to a solar region being assigned an
active region number by the NOAA Space Weather Prediction Center 
(http://www.swpc.noaa.gov) so that its 
evolution and activity can be tracked \cite{Gal2007}.
The formation and evolution of active regions are fundamental to
solar dynamic phenomena such as flares and Coronal Mass 
Ejections and their effect on the Earth environment.  

The visibility of sunspots is currently thought to be limited only
by geometrical effects arising from projection of the solar sphere 
onto a 2D image, effects referred to as foreshortening. In this
paper we present results obtained serendipitously while 
analysing sunspot data by means of Virtual Observatory tools,
showing that the visibility of small sunspots is 
much poorer than predicted by the foreshortening model.

\section{Data analysis}

We analysed sunspot group data from the USAF/Mount Wilson
catalogue, for times between 1 December 1981 to 31 December 2005,  
covering two and a half solar cycles.
We processed the data by means of AstroGrid workflows.
AstroGrid (http://www.astrogrid.org) \cite{Wal2006} is the UK's contribution to a global 
Virtual Observatory (VO), aimed at allowing seamless access to a variety
of astronomical data, and at providing efficient software
tools for data analysis. 
Our workflows processed all entries in the catalogue 
and extracted properties of individual regions,
identified by their NOAA region number, including the time and location of their first and
last observation.
The longitude of each region at 12:00UT on the days when it was first and last
observed was calculated.

\begin{figure} 
\begin{center}
\includegraphics[width=0.90\linewidth]{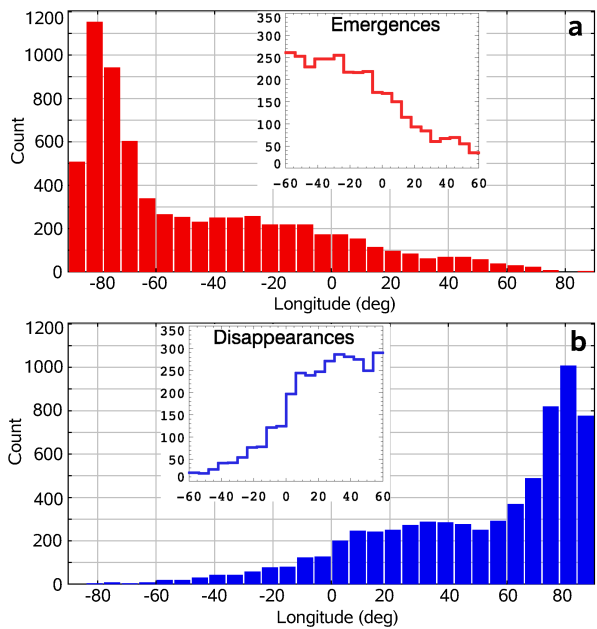}
\caption{Histograms of number of sunspot regions versus their longitude
at 12:00 UTC on the day when they were first (a) and last (b) observed . Each
longitude bin is 6$^{\circ}$. Data are from the USAF/Mt Wilson catalogue
of sunspot groups, for the time range 1 December 1981 to 31 December 2005. 
The total number of regions is 6862. Longitude 0$^{\circ}$ corresponds to the
Earth-Sun line, negative values to eastern longitudes and positive
values to western longitudes. Insets show the histograms for the range
[$-$60$^{\circ}$, $+$60$^{\circ}$].}
\end{center}
\end{figure}

Figure 1 shows histograms of the locations
at which sunspot regions were first and last observed,
with 0$^{\circ}$ longitude
corresponding to the Earth-Sun line. 
As the Sun rotates, many spots first come into
view near the east (or \lq rising\rq) limb, causing 
the peak to the left in Fig. 1-a.
Similarly, the peak in Fig. 1-b corresponds to 
regions that rotated out of view.
Regions first observed sufficiently far from the east limb,
are generally assumed to be \lq new\rq, indicating the emergence
of magnetic flux through the photosphere.  
As spots move in longitude by approximately 14.3$^{\circ}$ each day, regions
first observed to the west of $-$60$^{\circ}$ are typically
described as new emergences (see inset in Fig. 1-a).
Similarly regions last seen to the east of $+$60$^{\circ}$ are
interpreted as having decayed to the point that a spot is no
longer visible (see inset in Fig. 1-b).
 
The insets in Fig. 1 display strong east-west asymmetry.
A total of 825 new regions are seen to emerge in the bin 
[$-$60$^{\circ}$,$-$40$^{\circ}$], while only
177  in  [$+$40$^{\circ}$,$+$60$^{\circ}$], a ratio of 4.7:1.
What is the cause of the strong asymmetry in these curves? 
Why should the number of new regions emerging in any given longitude bin not
be constant? What is the true rate of magnetic flux emergence on the Sun?

An asymmetry in the location of emergence of new sunspots as viewed 
from Earth was discovered 100 years ago \cite{Mau1907}
and attracted the attention of famous physicists, who demonstrated that
a visibility function that favours 
observations in the centre of the disk, and the curve of
evolution of a spot's size, can produce such asymmetry \cite{Sch1911,Min1939}. 
The graphical representation introduced by Minnaert (1939) makes the cause
of the asymmetry immediately clear. However, Minnaert himself went on to
{\it assume} that the only factor limiting the visibility of sunspots is
geometrical effects associated with foreshortening. 
This results in a visibility function proportional to $\sec{\lambda}$, 
with $\lambda$ the longitude, and 
poor visibility only very near the limb.
The latter assumption has remained undisputed until the present day.
Moreover, appreciation of the cause of the east-west asymmetry appears to have
been lost, and 
asymmetries in sunspot parameters have since been ascribed, e.g., to 
observer bias or systematic inclination of the magnetic flux tubes
\cite{How1991}.

The large asymmetries in the insets of Fig. 1, however, demonstrate that 
the visibility of new sunspot formation in [$-$60$^{\circ}$, $+$60$^{\circ}$] is poor, 
and flux
{\em appears} to emerge and disperse at locations (and times) that are far from the
locations (and times) where {\em actual} emergence and decay took place.

We use the data of Fig. 1  
to derive the rate of region emergence, the average ratio between the slopes of 
growth and decay phases of region evolution
and to determine the visibility function.
Let $s(\lambda)$ be the visibility function, giving the minimum actual (as
opposed to apparent) area
that a sunspot region needs to reach to be visible at longitude $\lambda$. 
$s$ is expected to have a minimum at $\lambda$=0, where
visibility is best.
Assuming that, over a given time period, the number of magnetic flux
emergences in a unit longitude bin 
is a constant $N_1$, 
the number $N(\lambda)$ of regions {\em observed} emerging in a unit
bin at longitude $\lambda$ is \cite{Sch1911} [See also Appendix A]:
\begin{equation} 
N(\lambda)=N_1 \, \left[ 1 - \frac{\Omega}{k}\, s'(\lambda) \right]
\end{equation}
where $\Omega$ is the solar rotation rate and $k$ is the linear growth constant 
of a region's area with time.
Similarly, we obtain for the number $n(\lambda)$ of regions 
seen to disappear at $\lambda$: 
\begin{equation}
n(\lambda)=n_1 \, \left[ 1 + \frac{\Omega}{l}\, s'(\lambda) \right]
\end{equation}
with $n_1$ the number of regions that 
reach peak area in a unit longitude interval and $l$ the decay time constant, $l$$>$0. 

\begin{figure} 
\begin{center}
\hspace*{-1cm}
\includegraphics[width=0.95\linewidth]{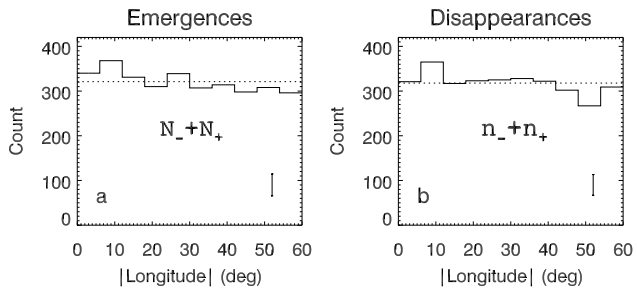}
\caption{Values of the sums of counts in positive
and negative longitude bins at either side of the Earth-Sun line,
as a function of absolute value of the longitude,
for region emergences (a) and disappearances (b). Dotted lines indicate
the mean. The sizes
of error bars are also shown. The size of each bin is 6$^{\circ}$.}
\end{center}
\end{figure}

Considering two bins centred at +$\lambda$ and
$-$$\lambda$ 
and indicating the number of regions seen emerging in them 
as $N_+$ and $N_-$ respectively, 
one can write Eq.~(1) for each of the two bins, and obtain
expressions for $N_+$ and $N_-$. By adding these together and
assuming that the visibility function is symmetric
with respect to $\lambda$=0, so that
$s'(-\lambda)$=$-$$s'(+\lambda)$, 
one finds that $N_+$+$N_-$=$2 \, N_1$ \cite{Sch1911}
[a misprint appears in Schuster's expression on his p. 319].
For regions disappearing in the same two bins,
from Eq.(2): $n_+ + n_-$=$2 \, n_1 $=$2 N_1$. Here we
assume that the average number $ n_1 $ of regions that
reach peak area in a unit longitude bin is equal to the average number $N_1$ 
of regions that emerge in a unit longitude bin.

Figure 2-a shows values of $N_+$+$N_-$ versus $|\lambda|$
obtained from the data of Fig. 1-a  by adding counts in positive
and negative $\lambda$-bins at either side of $\lambda$=0.
Figure 2-b shows $n_+$+$n_-$ obtained from the disappearences data of Fig. 1-b.
$N_+$+$N_-$ and $n_+$+$n_-$ are approximately constant, as predicted by
Schuster's theory, and we can use
their mean values (indicated by the dotted lines in Fig. 2) 
to calculate $N_1$, the actual number of sunspot
regions emerging in a unit bin. We
find $N_1$=160.55$\pm$11.41 from  Fig. 2-a and
158.95$\pm$12.19 from  Fig. 2-b.
Considering that all regions in the USAF/Mt Wilson catalogue are included
in our analysis,
and that the catalogue is compiled from observations 
made at discrete intervals of time,
the data fit Schuster's simple
theory remarkably well. There is excellent agreement in the values of $N_1$ 
derived separately from emergences and disappearances.
The data shown in Fig. 2 are consistent with a
constant value of $N_+$+$N_-$ and $n_+$+$n_-$ as would be expected 
from a visibility function that is symmetric with respect to $\lambda$=0.
Figure 2-a does display a small slope, while this is not seen in
Fig. 2-b. 
Several authors discussed the issue of whether the magnetic flux tubes of
active regions present a systematic inclination with respect to the direction
perpendicular to the solar surface. An analysis of magnetograms showed 
a systematic inclination of magnetic flux tubes of growing active regions 
of 24$^{\circ}$ in the W-E direction, i.e. trailing the rotation \cite{How1991}. 
This would result in better visibility of young regions in the west, the 
opposite of the effect we find.
We conclude that our data do not show evidence for a strong inclination
although this issue may need to be further investigated with other data.

From Eqs.(1) and (2), by solving for the
first derivative $s'(\lambda)$, and by dividing one equation by the
other,
we find an expression for the ratio
$k/l$ between the growth and decay constants characterising sunspot
region evolution.
We find a mean value $k/l$=1.37$\pm$0.26 (s.d.), from 
16 longitude bins. (Here, we exclude 4 longitude bins
near $\lambda$=0 in which $N/N_1$ and $n/n_1$ are close to 1, giving two 
terms close to zero to be divided by each other to find $k/l$).
Our value for $k/l$ implies that, on average,
the duration of the decay phase of an active region
is only 1.37 times larger than that of the rise phase. 
This is very different from previous estimates,
according to which active regions reach maximum
development very quickly, e.g. within 5 days of emergence \cite{Har1993},
and decay slowly.
The ratio being different from 1 is the cause of the lack of
complete symmetry in the distributions of emergences and disappearances
(see insets of Fig. 1).
While in general the curve of evolution of sunspot regions will be 
complex, its linear envelope, describing a triangle with slopes $k$ and $l$,
is a useful first approximation.
By means of modelling, we found that
the asymmetry in [$-$60$^{\circ}$, $+$60$^{\circ}$] is largely independent
of the actual total lifetime of regions, while lifetime is the
key factor influencing the height of the peak in Fig. 1-a.

\begin{figure} 
\begin{center}
\hspace*{-1cm}
\includegraphics[width=0.8\linewidth]{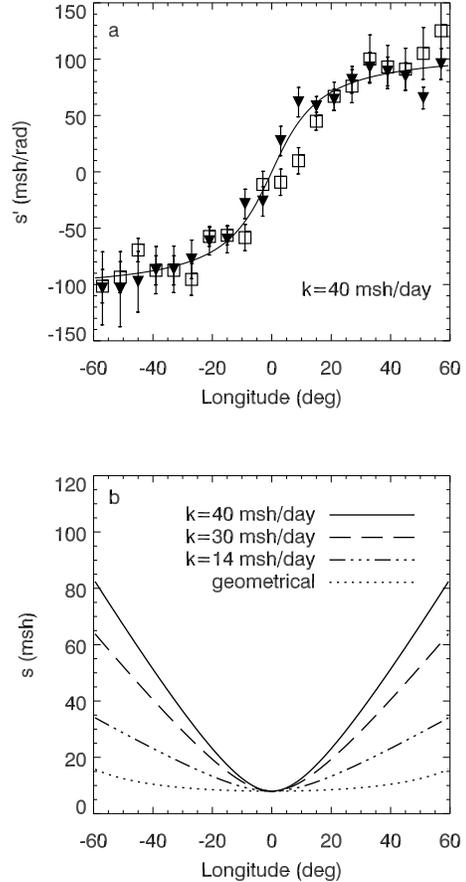}
\caption{(a)
Derivative of the visibility function from emergence (squares)
and disappearance (triangles) data, for  k=40 msh/day and $k$/$l$=1.37, 
with msh=millionths of the visible solar hemisphere; the solid line shows the
best fit to the data points.
(b) Visibility function $s(\lambda)$, giving the minimum area  
a sunspot needs to reach to be detected at longitude $\lambda$, for several
values of $k$. The dotted line 
shows the function $A_{min}\,\sec{\lambda}$.}
\end{center}
\end{figure}

Having obtained $N_1$ and the ratio $k/l$, values of
the derivative of the visibility function, $s'$, given by Eqs.(1) and (2), 
depend on a single parameter, the growth constant $k$. 
In Fig. 3-a we plot $s'$ data points, calculated 
using Eqs.(1) ({\it squares}) and (2) ({\it triangles}) for a value of the growth constant $k$=40 msh/day
and using $k/l$=1.37 (msh indicates millionths of the visible solar hemisphere).
We fit the  function 
$s'(\lambda)$= $c_1\, \tan^{-1}{(c_2\, \lambda)}$ to the data 
and obtain a best fit when $c_1$=117.0 and $c_2$=4.7 (the solid
line in Fig. 3-a). By integration we
obtain $s(\lambda)$=$c_1/c_2 \,[x \tan^{-1}{x} -1/2 \ln(1+x^2)]+A_{min}$,
where $x$=$c_2\, \lambda$ (with $\lambda$ in radians) and
$A_{min}$ is the minimum of the visibility function, set equal to 8 msh,
the approximate minimum area required at the disk centre for a group to be included
in the catalogue.
Figure 3-b shows $s(\lambda)$ for the fit of Fig. 3-a (solid line)
and for those obtained for two other values of k.
The dotted line shows $A_{min}\, \sec{\lambda}$, the visibility function
expected from geometrical considerations describing foreshortening.

Sunspot regions are in reality characterised by a distribution of
growth and decay constants. The curves of Fig. 3-b therefore need
to be interpreted as the visibility functions arising when the most
probable value of $k$ in the distribution is equal to the numerical
value given. Recurrent sunspots were found to have a decay rate constant
$l$$\approx$10 msh/day \cite{Mar1993}, giving $k$=14 msh/day using our $k/l$ ratio. Due
to the very slow decay of recurrent spots the latter is a
likely good lower limit for $l$ and consequently $k$.
Even when $k$=14 msh/day, sunspot visibility
is much worse than expected from projection effects only.
To recover the geometrical visibility curve from the data, it would be
necessary to assume $k$=3 msh/day, a value not consistent with
observations.

\section{Discussion}

Figure 3-b demonstrates that the visibility of small sunspots
is much poorer than expected from geometrical effects
associated with foreshortening.
It shows that the minimum area required for a spot to be detected 
at $\lambda$=$\pm$30$^{\circ}$ is more than twice the threshold area 
at  $\lambda$=0$^{\circ}$ if $k$=14 msh/day and almost 4 times if 
$k$=30 msh/day.  The centre-to-limb variation in visibility is remarkably
large.

We investigated whether the asymmetries shown in Fig. 1 display any solar
cycle dependence, and found no evidence of it.
The USAF/Mount Wilson locations and times of first appearances
agree with those from SOHO/MDI continuum data (as verified manually for a 
sample of regions). It is known that seeing associated with ground-based data does not
cause significant reduction of visibility \cite{Gyo2004}.

How many spots are affected by the visibility effects here described?
The distribution of sunspot areas 
measured at a single longitudinal location on the solar disk is lognormal
\cite{Bog1988}
and the number of spots with area at Central Meridian around 10 msh is more than 
2 orders of magnitude larger
than the number of spots with area of about 100 msh.
Therefore the majority of sunspots will cross the visibility curve shown in 
Fig. 3-b.

Our results have a number of important implications.
The first is that the radiative processes that make
a small region of strong magnetic field appear as a dark spot,
have a strong centre-to-limb variation.
This may prove important 
for the study of sunspots' 3D structure and will require further
investigation.
Whether larger spots are also affected by the same process will also need
further
study. Reports of centre-to-limb variations of corrected 
sunspot areas have appeared in the literature \cite{Gyo2004, Hoy1983}.
Faculae have a large centre-to-limb variation, and their 
contrast changes sign as one moves towards the disk centre, resulting in
their being darker than the surrounding photosphere at the disk centre \cite{Law1993}.
This demonstrates that the appearance of photospheric magnetic flux tubes
strongly depends on the viewing angle.

The second implication  is that actual distributions of sunspot lifetimes and areas
may differ from the apparent ones derived from observations.
The latter have been used to constrain mechanisms of sunspot formation and decay
\cite{Sol2003, Pet1997, Mar1993}.
A large number of regions reported  of
short duration
may in fact have longer lifetimes, and be crossing
in and out of the visibility curve. 
The Gnevyshev-Waldmeier law, stating that a sunspot's lifetime
increases linearly with its maximum size,
may need to be reassessed in light of our results.
Poor visibility of region emergence means that the actual
time of magnetic flux emergence can be much earlier than the apparent
time, e.g.~for a region seen to emerge at $\lambda$=$-$50$^{\circ}$,
by approximately 2 days. On the other hand, a large fraction of new emergences 
in the western portion of the solar disk go undetected. By using the data
in the inset of Fig.~1-a for $\lambda$$>$0 and the value of actual number
of emergences $N_1$=160 obtained from the data of Fig. 2, 
we obtain that 44\% of new spots emerging in [0$^{\circ}$, $+$60$^{\circ}$]
were invisible.

The presence of a sunspot is key to a solar active region 
being assigned an Active Region Number by the NOAA Space Weather Prediction 
Center
(see e.g. Dalla et al. (2007) for the full list of criteria).
A region that has been given a NOAA number is monitored and its activity tracked
\cite{Gal2007}. New regions emerging in the west of the Sun with
their spots being invisible are missed and not tracked.
We conclude that current criteria for assigning Active Region Numbers
may need to be revised and that
EUV solar images may need to be routinely used to supplement white-light
information.

Sunspots are well known to cause depletions in the total solar irradiance
(TSI) \cite{Fou2006} and many models of TSI variation need as input information
on the number and areas of sunspots. 
While the sunspots that most affect TSI are the largest ones, of area typically
well above the visibility threshold shown in Fig. 3-b, 
our results impact TSI studies because they demonstrate that the
apparent age and stage of development of a sunspot may not
correspond to the actual ones. The latter information is required
when studying the time dependence of sunspot effects on TSI and
whether the age of a region is an important factor in determining
the magnitude of TSI decrease.

The asymmetry in the distribution of emergences was obtained, 
unexpectedly, during a study aiming at cross correlating catalogues of
sunspot regions and flares, by means of AstroGrid workflows \cite{Dal2007}. 
This demonstrates the usefulness of VO tools in making new science 
possible, by provision of better
tools for analysis of large datasets. 

\section*{Acknowledgements}
This research has made use of data obtained using, or software provided
by, the UK's AstroGrid Virtual Observatory Project, which is funded by the
Science and Technology Facilities Council and through the EU's
Framework 6 programme. 
We thank Dr Frank Hill for pointing out that an east-west asymmetry
had been reported in the literature and Dr Hugh Hudson for comments.
L.F. acknowledges the support of PPARC Rolling Grant
PP/C000234/1 and financial support by the European Commission through the SOLAIRE Network
(MTRN-CT-2006-035484).

\Online

\begin{appendix} 
\section{Schuster's equation and Minnaert's graphical representation}

The fact that a visibility function favouring sunspot observations
in the centre of the solar disk, should result in an east-west
asymmetry in the number of regions seen emerging at the Sun, is
not immediately intuitive. In this Appendix we summarise Schuster's
derivation of Eq.(1) of our paper (Schuster 1911) and describe 
Minnaert's graphical 
representation, from which the cause of the asymmetry becomes immediately
apparent (Minnaert 1939).

\begin{figure}[b]
\centering
\includegraphics[width=0.65\linewidth]{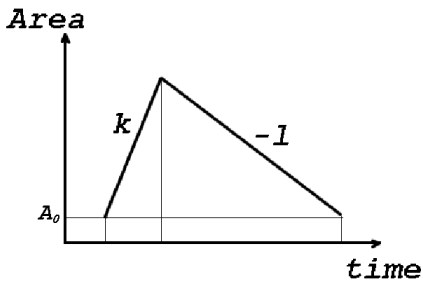}
\caption{Schematic of time evolution of a sunspot's area. Here $A_0$ is the
  area of the spot at the time when it first emerges through the photosphere.}
\label{appfig1}
\end{figure}

Two phenomena combine to produce the effect here described. The first is the
fact that sunspot regions are evolving: their evolution can be characterised
in terms of their area and in a zero-th order approximation can 
be described by a curve such as shown in Fig.~\ref{appfig1}:
a growth phase with slope $k$ and a decay with slope $-$$l$ ($l$$>$0).
The second phenomenon is the solar rotation, which carries sunspots which 
emerged in the east of the Sun towards regions of better visibility
(i.e.~towards the centre of the disk) and regions that emerge in the west 
towards regions of worse visibility.

\begin{figure}[t]
\centering
\includegraphics[width=0.7\linewidth]{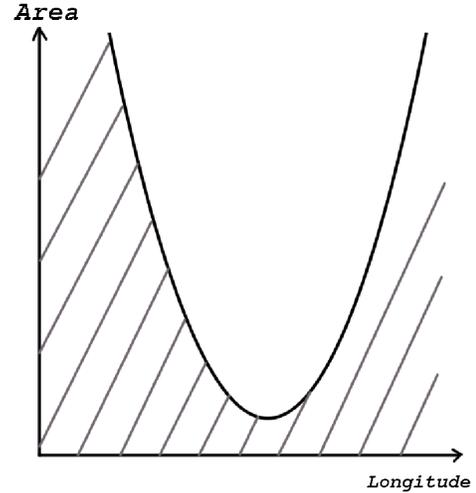}
\caption{Schematic of Minnaert's graphical representation (adapted from
  Minnnaert 1939).
  The black curve is the visibility function $s(\lambda)$ and the grey parallel
  curves represent the growth curves of sunspot regions.}
\label{appfig2}
\end{figure}
The combination of region evolution and rotation, together with
the specific form of the visibility function, determines
whether a sunspot region is seen or not, and the location and
time of its first appearance (and of its disappearance)
to an Earth observer.
This is clear from Minnaert's graphical representation 
(Minnaert 1939), as shown in Fig.~\ref{appfig2}.
The curve in the diagram represents the visibility
function $s(\lambda)$, giving the actual area that a sunspot
region needs to reach to be visible at longitude $\lambda$.
The parallel lines at a slope represent growth phases of sunspots.
When a sunspot region growing along a given  line crosses
$s(\lambda)$, it becomes visible.
From this graphical representation it can be seen that if 
the visibility function has a strong centre-to-limb variation,
many spots forming in the west are invisible. An east-west
asymmetry in the number of regions observed emerging thus results,
and the asymmetry depends on the gradient $s'$ of the
visibility function. 

The process qualitatively represented by Minnaert's graph 
was quantitatively described by Schuster (1911), as follows.
Let $\lambda_1$ indicate the longitude at which a sunspot forms,
and $\lambda$ the longitude at which it is first seen because it
has crossed the visibility curve $s(\lambda)$. 
The two longitudes are related by the following equation:
\begin{equation}\label{eq.app1}
A_0+ k \,\, \frac{\lambda -\lambda_1}{\Omega} = s(\lambda)
\end{equation}
where $A_0$ is the area of the spot when it first emerges, $k$ is the 
slope of the growth phase and $\Omega$ the solar rotation rate. 
Eq. \ref{eq.app1} expresses the fact that the area at the location where the
spot is first seen is equal to $s(\lambda)$.

Eq. \ref{eq.app1} can be re-arranged to give:
\begin{equation}\label{eq.app2}
\lambda -\frac{\Omega}{k} \, s(\lambda) = \lambda_1 -\frac{\Omega}{k} \, A_0
\end{equation}
which, differentiated, gives:
\begin{equation}\label{eq.app3}
d \lambda \left[ 1 -\frac{\Omega}{k} \, s'(\lambda)\right] = d \lambda_1
\end{equation}

If $N_1$ indicates the number of sunspots emerging in a unit longitude bin 
and $N(\lambda)$  the number of regions {\em observed} emerging in a unit
bin at longitude $\lambda$, then $N_1 \,d \lambda_1 $=$N(\lambda)\, d
\lambda$, which gives, using Eq. \ref{eq.app3}:
\begin{equation} 
N(\lambda)=N_1 \, \left[ 1 - \frac{\Omega}{k}\, s'(\lambda) \right]
\end{equation} (Schuster 1911) (Eq (1) of our paper).

Applying the same derivation to the decay phase, characterised
by a slope $-$$l$, we obtained Eq. (2) of our paper.

\end{appendix}
\end{document}